\documentclass[12pt]{article}
\usepackage{youngtab}
\def\beq{\begin{equation}}
\def\eeq{\end{equation}}
\def\bea{\begin{eqnarray}}
\def\eea{\end{eqnarray}}
\def\bq{\begin{quote}}
\def\eq{\end{quote}}

\def\gappeq{\mathrel{\rlap {\raise.5ex\hbox{$>$}}
{\lower.5ex\hbox{$\sim$}}}}

\def\lappeq{\mathrel{\rlap{\raise.5ex\hbox{$<$}}
{\lower.5ex\hbox{$\sim$}}}}

\def\Toprel#1\over#2{\mathrel{\mathop{#2}\limits^{#1}}}

\begin{document}
\pagestyle{empty}
\begin{flushright}
{CERN-TH/2003-316}\\
{DSF 43/03}
{LAPTH-1018/03}\\
\end{flushright}
\vspace*{5mm}
\begin{center}
{\bf NARROW WIDTH PENTAQUARKS}
\\
\vspace*{0.5cm}
{\bf F. Buccella} \footnote{On leave of absence from Dipartimento di
Scienze Fisiche, Universit\`{a} di Napoli
 ``Federico II", Complesso Universitario di Monte S. Angelo, Via
Cintia,
I-80126 Napoli, Italy, and INFN, Sezione di
Napoli}\\
\vspace{0.1cm}
and\\
\vspace{0.1cm}
{\bf P. Sorba} \footnote{On leave of absence from
LAPTH, 9 chemin de Bellevue, BP 110,
F - 74941 Annecy-le-Vieux Cedex, France}\\
\vspace{0.8cm}
Theoretical Physics Division, CERN\\
CH - 1211 Geneva 23, Switzerland\\

\vspace*{1.5cm}

{\bf ABSTRACT} \\ \end{center}

\vspace*{2mm}

A general study of pentaquarks built with four quarks in a $L =
1$ state and an antiquark in  $S$-wave shows that several of such
states are forbidden by a selection rule, which holds in the limit
of flavour symmetry, to decay into a baryon and a meson final
state.  We identify the most promising $\overline{10}$ multiplet
for the classification of the $\Theta^+$ and $\Xi^{--}$ particles
recently discovered  with the prediction of a narrow width for
both of them.

\vspace*{1.3cm}

\begin{flushleft} CERN-TH/2003-316 \\
{DSF-43/03}\\ {LAPTH-1018/03}\\
December 2003
\end{flushleft}
\newpage

\setcounter{page}{1}
\pagestyle{plain}

A narrow KN resonance, called $\Theta^+$, has been found at 1540 MeV in
different experiments \cite{N}: it seems to
have $I = 0$, since no similar state has been found in $K^+p$
scattering.\\
More recently, at CERN, a $\Xi^{--}$ state and a $\Xi^{0}$ state
have been found at a mass of 1862 MeV with a width below the
detector resolution of 18 MeV \cite{A}.  Since a $\Xi^{+}$ is
expected to exist on the basis of isospin invariance, one may
conclude that all the exotic states (impossible to build with
three quarks) of a flavour $SU(3)_F$ $\overline{10}$
representation have been found.  Interestingly enough, a
$\overline{10}$ of $SU(3)_F$ with $J^p = 1/2^+$ is predicted in
the framework of the Skyrme model \cite{S}, in the same group of
states of the better-established $(8,1/2)^+$ and $(10,3/2)^+$
traditionally classified in the 56-dimensional representation of
flavour-spin $SU(6)_{FS}$ \cite{GR}, and the value of the mass of
the $Y = 2, I = 0$ state happened to be predicted at the right
value \cite{DPP}. In fact, one of the authors (D.D.) has been very
active in promoting the experimental search for that state. These
states can be thought to be pentaquarks, consisting of four quarks
and one antiquark, with $\Theta^+$ being a $uu dd \bar{s}$ state
and the $\Xi^{--}$ resonance  a $dd ss \bar{u}$ state. Hereafter,
we propose to construct the wave functions of $4q$ states,
consistent with the total antisymmetry dictated by Fermi-Dirac
statistics, in colour and other degrees of freedom. In order to
constitute a $SU(3)$ colour singlet together with the antiquark,
the four-quark state should transform as a $3$ of $SU(3)_c$,
corresponding to the Young tableau [2,$1^2$]. Therefore, to build
a complete antisymmetric wave function, the symmetry prescription
with respect to the other variables of the four quarks should
correspond to the Young tableau [3,1]. Should their spatial wave
function be totally symmetric in the absence of an orbital angular
momentum $(\vec{L} = 0)$, they have to transform as the $210$
representation of flavour spin $SU(6)_{FS}$, which has the
symmetry properties of the just-mentioned Young tableau.  The
corresponding pentaquark would then be constructed by composing
the $SU(3)_F {\times} SU(2)_S$ states of the $210$ with the
$\overline{6} = (\overline{3},2)$ antiquark and
 with the orbital momentum of the $\bar{q}$
with respect to the four-quark system: if this last one is zero --
$\bar{q}$ in a $S$-wave -- one should get negative parity states.
To get positive parity states with an $S$-wave $\bar{q}$, one
could consider, as in \cite{JW} \cite{SR}  , $L = 1$ four-quark
states. To this extent we write the identity: \bea \sum^{4}_{i=1}
\; \vec{r}_i \wedge \vec{p}_i & = & \frac{1}{4} \left \{ \left
(\sum^{4}_{i=1} r_i \right ) \wedge \left ( \sum^{4}_{i=1} p_i
\right ) + ( \vec{r}_1 + \vec{r}_2 - \vec{r}_3 - \vec{r}_4 )
\wedge ( \vec{p}_1 + \vec{p}_2 - \vec{p}_3 - \vec{p}_4 ) \right.
\nonumber \\
& + & \left. ( \vec{r}_1 + \vec{r}_3 - \vec{r}_2 - \vec{r}_4 ) \wedge (
\vec{p}_1 + \vec{p}_3 - \vec{p}_2 - \vec{p}_4 )
\right. \nonumber \\
& + & \left. ( \vec{r}_1 + \vec{r}_4 - \vec{r}_2 - \vec{r}_3 )
\wedge ( \vec{p}_1 + \vec{p}_4 - \vec{p}_2 - \vec{p}_3 ) \right \}
\label{1} \eea where $\vec{r}_i$ and $\vec{p}_i$ stand for the
coordinates (position and momentum) of the i-th quark.  The first
term on the right-hand side of Eq. (1) is the centre-of-mass
angular momentum, while the three other terms are the relative
angular momenta, commuting between each other.  Their
corresponding eigenfunctions have to be combined with the
$SU(6)_{FS}$ functions in order to satisfy a global symmetry
conforming to the [3,1] Young tableau.  Thus, composing with the
color $3$ wave function part, one will finally obtain a completely
antisymmetric wave function for the $4q$ state. For convenience,
let us introduce the notation: \beq \vec{r}_{1i} = \frac{1}{2}
(\vec{r}_1 + \vec{r}_i - \vec{r}_j - \vec{r}_k ) \label{2} \eeq
with $i,j,k = 2,3,4$ all different. Considering as an example the
exchange $\vec{r}_1 \leftrightarrow \vec{r}_2$, one notes that
$\vec{r}_{12}$ is unchanged, while $\vec{r}_{13} \leftrightarrow
-\vec{r}_{14}$, implying analogous transformations on the
spherical harmonics:  $Y_{1m} (\vec{r}_{13}) \leftrightarrow -
Y_{1m} (\vec{r}_{14})$. Now, there are four different $SU(6)$
irreducible representations which are present in the decomposition
of the product $6 {\times} 6 {\times} 6 {\times} 6$, the
representations $126, 210, 105$ and $105'$ with corresponding
Young tableaux $[4], [31], [2^2]$ and $[21^2]$ respectively, for
which one can determine the totally antisymmetric wave function
under $SU(6)_{FS} {\times} O(3)_L {\times} SU(3)_C$. The result is
as follows:
\beq Ant. \left \{ F_{126} (\vec{r}_i \cdot \vec{r}_j
) \, \psi^{126}_{ \{A,B,C,D\} } \,[Y_{1m} (\vec{r}_{12}) + Y_{1m}
(\vec{r}_{13}) + Y_{1m} (\vec{r}_{14}) ] \; f_{[\beta , \gamma ,
\delta] \alpha} \right \} \label{3} \eeq \bea & Ant. & \left \{
F^{210} (\vec{r}_i \cdot \vec{r}_j)  \, \left [ (\psi^{210}_{ \{
A,C ,D\} ,B} + \psi^{210}_{ \{ A,B,D \} ,C}) ( Y_{1m}
(\vec{r}_{12}) + Y_{1m} (\vec{r}_{13}) ) \right. \right.
\nonumber \\
& + & \left ( \psi^{210}_{\{A,B,C\},D} + \psi^{210}_{\{A,C,D\},B}
\right ) \left ( Y_{1m} (\vec{r}_{12}) + Y_{1m} (\vec{r}_{14})
\right )
\nonumber \\
& + &  \left ( \psi^{210}_{\{A,B,C\},D} + \psi^{210}_{\{A,B,D\},C}
\right ) \left. \left. \left ( Y_{1m} (\vec{r}_{13}) + Y_{1m}
(\vec{r}_{14}) \right ) \right ] \, f_{[\beta , \gamma , \delta ],
\alpha} \right\} \label{4} \eea \bea & Ant. & \left \{  F^{105}
(\vec{r}_i \cdot \vec{r}_j)  \, \left [ \psi^{105}_{ \{ A,B\}
\{C,D \} }
Y_{1m} (\vec{r}_{12})  \right. \right. \nonumber \\
& + &  \psi^{105}_{ \{ A,C \}  \{ B,D \} } \,  Y_{1m}
(\vec{r}_{13})
\nonumber \\
& + & \left. \left. \psi^{105}_{ \{ A,D \} \{ B,C \} } \,  Y_{1m}
(\vec{r}_{14}) \right ] f_{[\beta , \gamma , \delta ], \alpha}
\right \} \label{5} \eea \bea & Ant. & \left \{  F^{105'}
(\vec{r}_i \cdot \vec{r}_j)  \, \left [ \psi^{105'}_{ \{ B,C
\},D,A} Y_{1m} (\vec{r}_{14})  \right. \right.
\nonumber \\
& + &  \psi^{105'}_{ \{ B,D \} , C,A } \,  Y_{1m} (\vec{r}_{13})) \nonumber \\
& + & \left. \left. \psi^{105'}_{ \{ C,D \},  B,A  } \,  Y_{1m}
(\vec{r}_{12}) \right ] f_{[\beta , \gamma , \delta ], \alpha}
\right \} \label{6} \eea where the $SU(6)$ space and colour
coordinates of the four quarks are denoted by $(A, \vec{r}_1 ,
\alpha ; B , \vec{r}_2,$ $\beta ; C, \vec{r}_3 , \gamma , D ,
\vec{r}_4 , \delta )$, the $F$'s are scalar functions in the
variables $\vec{r}_i \cdot \vec{r}_j$ including a renormalization
factor, the $Y_{1,m}$ are the spherical harmonics and the $\psi$
are defined on $SU(6)$ flavour-spin variables with the symmetry in
the lower indices corresponding to the Young tableaux: \beq \{
A,B,C,D \} \quad \rightarrow  \quad \young(ABCD) \label{7} \eeq
\beq \{  A,B,C \} ,D \quad \rightarrow \quad \young(ABC,D)
\label{8} \eeq \beq \{  A,B \} \{ C,D \} \quad \rightarrow \quad
\young(AB,CD) \label{9} \eeq \beq \{  A,B \},C,D \quad \rightarrow
\quad \young(AB,C,D) \label{10} \eeq respectively, where we first
symmetrize on the rows and then antisymmetrize on the columns. The
colour wave function $f_{[\beta , \gamma , \delta ] \alpha}$
corresponds to the Young tableau: \beq
\young(\beta\alpha,\gamma,\delta) \label{11} \eeq where we first
antisymmetrize on the column and afterwards symmetrize on the row.
In such a way the expressions within the $\{ \}$ are antisymmetric
with respect to the exchange of the three sets of variables $(B,
\vec{r}_2 , \beta ; C, \vec{r}_3 , \gamma ; D, \vec{r}_3 , \delta
)$.  Finally, the operator $Ant.$ has the task of antisymmetrizing
in the four variables.  It acts in the following way, transforming
the product
\begin{center}
\young(bcd,a)
\qquad
\young(\beta\alpha,\gamma,\delta)
\end{center}
which is antisymmetric in the variables $(b, \beta ; c,\gamma ; d,
\delta )$ into the combination: \bea & \young(bcd,a) & \quad
\young(\beta\alpha,\gamma,\delta) + \quad \young(acd,b) \quad
\young(\alpha\beta,\delta,\gamma)
\nonumber \\
& + & \young(abd,c) \quad \young(\alpha\gamma,\beta,\delta) +
\quad \young(abc,d) \quad \young(\alpha\delta,\gamma,\beta)
\label{12} \eea which is antisymmetric in the variables $(a,
\alpha ; b, \beta ; c, \gamma ; d, \delta )$. Note that in this
last operation we have made the identification $(a,b,c,d) \equiv
(A, \vec{r}_1 ; B, \vec{r}_2 ; C, \vec{r}_3 ; D, \vec{r}_4)$.\\
The decomposition of the different $SU(6)_{FS}$ representations
into $SU(3)_F {\times} SU(2)_S$ for $L = 1$ $4q$-states is
reported in Table 1, together with the resulting $SU(3)_F {\times}
SU(2)_J$ representations obtained by composing them with the
$(\bar{3},\frac{1}{2})$ and the orbital momentum.

\begin{center}
\begin{tabular}{|c|c|c|}\hline
4q $SU(6)_{FS}$ & $SU(3)_F$x$SU(2)_S$ & $4q\bar{q}$ L=1
$SU(3)_F$x$SU(2)_J$
\\
\hline

126=[4]     &    (15'=[4],2)    &      $(35+10 ,\frac{7}{2}+ 2
                                     \frac{5}{2}+
                                     2 \frac{3}{2} +\frac{1}{2})$
\\
            &    (15=[3,1],1)   &      $(27 + 10 + 8 , \frac{5}{2} +
                                     2 \frac{3}{2} + 2 \frac{1}{2}$)
\\
            &    ($\bar{6}$,0)  &      ($\bar{10} + 8 , \frac{3}{2} +
                                     \frac{1}{2}$)
\\
\hline
210 = [3,1] &    (15',1)        &      (35 + 10 , $\frac{5}{2} +
                                     2 \frac{3}{2} + 2 \frac{1}{2}$)
\\
            &    (15 , 2 + 1 + 0) &    (27 + 10 + 8 , $\frac{7}{2} +
                                     3 \frac{5}{2} + 5 \frac{3}{2} +
                                     4 \frac{1}{2}$)
\\
            &    ($\bar{6}$ , 1 ) &    ($\bar{10} + 8 , \frac{5}{2} +
                                     2 \frac{3}{2} + 2 \frac{1}{2}$)
\\
            &    (3, 1 + 0 )      &    (8 + 1 , $\frac{5}{2} +
                                     3 \frac{3}{2} + 3 \frac{1}{2}$)
\\
\hline
105=[$2^2$] &    (15', 1 )        &    ( 35 + 10 , $\frac{5}{2} +
                                     2 \frac{3}{2} + 2 \frac{1}{2}$)
\\
            &    (15 , 0 )        &    (27 + 10 + 8 , $\frac{3}{2} +
                                     \frac{1}{2}$)
\\
            &    ($\bar{6}$, 2 + 0 ) & ($\bar{10} + 8, \frac{7}{2} +
                                     2 \frac{5}{2} + 3 \frac{3}{2} +
                                     2 \frac{1}{2}$)
\\
            &    (3, 1 )        &      ( 8 + 1 , $\frac{5}{2} +
                                     2 \frac{3}{2} + 2 \frac{1}{2}$)

\\
\hline
105'=[2,$1^2$] & ( 15 , 1 + 0 )  &  ( 27 + 10 + 8 , $\frac{5}{2} +
                                     3 \frac{3}{2} + 3 \frac{1}{2}$)
\\
              &  ( $\bar{6}$ , 1 ) &   ($\bar{10} + 8 , \frac{5}{2} +
                                     2 \frac{3}{2} + 2 \frac{1}{2}$)
\\
              &  ( 3 , 2 + 1 + 0 ) &   ( 8 + 1 , $ \frac{7}{2} +
                                     3 \frac{5}{2} + 5 \frac{3}{2} +
                                     4 \frac{1}{2} $)
\\
\hline
\end{tabular}
\end{center}
\vspace{0.1cm}
\begin{center}
TABLE 1 : Representations of $(4q)$ states and $L = 1 (4q+\bar{q})$
states. For convenience, the $SU(2)$ representations are not denoted
by their dimensions - which is the case for their $SU(3)$ partners
- but by their highest weight.
\end{center}

Let us add that a state constructed with four $S$-wave quarks and
one antiquark with relative angular momentum $L = 1$ is also in
the $210$ representation of $SU(6)_{FS}$.  It therefore belongs to
one of its corresponding $SU(3)_F {\times} SU(2)_J$ multiplets
expressed in Table 1, but, of course, does not share the same wave
function as the one given by Eq.(4). All the four $SU(6)_{FS}, L =
1$ representations considered in Table 1 contain a
$(\overline{10}, \frac{1}{2} )$ multiplet of $SU(3)_F {\times}
SU(2)_S$.  However, the two last cases, i.e., the $105, L = 1$ ,
which corresponds to the proper four-quark antisymmetrization of
the two diquark model proposed in
 \cite{JW}, and the $105', L = 1$, share
a property \cite{SR} which strongly favours them for the
classification of the experimentally found $\Theta^+$ and
$\Xi^{--}$ states. Indeed, when the $\bar{q}$ picks up one of the
$q$ to build a meson -- in the case of $\Theta^+$,  the $\bar{s}$
together with a $u$ or a $d$ forms a $K$ -- two of the remaining
three $q$'s remain in  a $SU(6)_{FS}$ antisymmetric state, so that
this wave function is orthogonal to that of the totally symmetric
56 $SU(6)_{FS}$ representation: this implies a
\underline{selection rule} against the decay of the pentaquark
into a meson plus a baryon of the $1/2^+$ octet or of the $3/2^+$
decuplet (let us recall the $SU(3) {\times} SU(2)$ decompositions
of the $56 = (8, 1/2) + (10, 3/2)$.) This property seems appealing
to account for the narrowness of the discovered pentaquark states.
Since the selection rule has been found in the limit of flavour
symmetry, we expect it to be violated by $SU(2)_I$-breaking terms
for $\Theta^+ \to KN$ (as happens for $\eta \to 3 \pi$ decay) and
by $SU(3)_F$ breaking for $\Xi^{--} \to \Xi^- \pi^-$. There are
many $\bar{6}$ $SU(3)_F$ representations among the $105' + 105 \,
L = 1$ states which may build the $\overline{10}$ together with an
$S$-wave $\bar{q}$, where we wish to classify the $\Theta^+$ and
the $\Xi^{I=3/2}$ states. To choose between them, one should get
an even qualitative idea about the spectrum of these pentaquark
states with the twofold motivation of finding a sufficiently low
value for the mass of the $\Theta^+$, and further,
with small components along the $126 + 210, L = 1$.\\
The mass differences within the S wave $3q$ and $q\bar{q}$ states
are well reproduced by the QCD chromo-magnetic interaction
\cite{DGG}. The fact that the binding energies are given by
combinations of the Casimir colour-spin $SU(6)_{CS}$, $SU(2)_S$
and $SU(3)_C$ \cite{HS}, allows us, as in the Arima and Iachello
model for nuclear physics \cite{AI}, to deduce the general
properties of the pentaquark spectrum from group theoretical
considerations. The two-quark interaction is attractive if their
wave function is symmetric in colour and spin and repulsive in the
other case.  So the form of the Young tableaux for $SU(6)_{CS}$
dictates the shape of the spectrum: for the states of the $56$,
the ($10 \; 3/2^+$), which transforms as the totally antisymmetric
$20$ representation of $SU(6)_{CS}$, is heavier than the ($8 \;
1/2^+$), which transforms as the $70$ of $SU(6)_{CS}$, which has a
mixed symmetry.  For the negative parity states of the $70 \, L =
1$, one can explain the differences between the mass of the $J^P$
= $\frac{5}{2}$ resonances  and the mean values of the masses of
the $J^P$ = $\frac{3}{2}$ and $\frac{1}{2}$ resonances with a $LS$
coupling with a coefficient of $40MeV$ and no contribution from
the chromo-magnetic quark interaction ( we would rather expect a
$\frac{1}{3}$ screening factor from the orbital angular momentum
rather than a complete one),which would follow from the fact that
the states constructed with the ($8, S = 3/2$) transform as a $20$
of $SU(6)_{CS}$ and  the states of the $(10 + 8 + 1, S = 1/2)$
transform as a $70$ under $SU(6)_{CS}$. Nevertheless, even in
absence of an effect of the spin forces for the $ L = 1$ states of
$4q$ we are considering , it is useful to classify the $SU(3)_F
{\times} SU(2)_S$ multiplets listed in Table 1 according to their
transformation properties with respect to $SU(6)_{CS}$ in order to
simplify the diagonalization of their interaction with the
$\bar{q}$. As in the case of $3q$ states, we find a spectrum,
factorized in flavour and colour, as described in Table 2:
\begin{center}
\begin{tabular}{|c|c|}\hline
$SU(6)_{CS}$ & $SU(3)_F {\times} SU(2)_S$ \\ \hline
210 & (15 + $\bar{6}$ + 3, S = 1 + 0) \\ \hline
105 & (15 + 3, S = 1) \\ \hline
105' & (15' + 15 + $\bar{6}$ + 3, S = 2 + 1 + 0) \\ \hline
$\overline{15}$ & (15, S = 1) \\ \hline
\end{tabular}
\end{center}
\vspace*{0.1cm}
\begin{center}
Table 2: Colour spin versus flavour and spin for $L = 1$ colour triplet
$4q$ states.
\end{center}

For the interested reader, we summarize how we have constructed
Table 2.\\
We selected in Table 1 the $SU(3)$ $3$ representations (in this
case we think of colour, but group theory does not care about the
interpretation) and we consider also the $\overline{15}$ of
$SU(6)$, which contains a $(3, S = 1)$ multiplet. By considering
the spin associated to the $3$ of $SU(3)$ of each $SU(6)_{CS}$
representation, we find its spin content. To find the flavour
content, we compose the Young tableau associated to each
$SU(6)_{CS}$ representation with the Young tableau [3,1]
corresponding to the orbital momentum, and we consider the dual
tableaux of the ones obtained (by dual of a Young tableau, we mean
that obtained by composing with the diagram $[1^4]$).\\
To understand the form of the spectrum of the pentaquark states
that can be built with the $L = 1$ $4q$ and a $\bar{q}$ in an
$S$-wave with respect to them, it is useful to make an analogy
with $q \bar{q}$ states, showing the consequences of the QCD
chromo-magnetic force for them. In fact, this force is strongly
attractive \cite{DGG} for $SU(6)$ singlets, which is the case of
the pseudoscalar mesons, and slightly repulsive for the colour
singlet spin-1 states, i.e., the vector mesons, which are
classified in a $35$ $SU(6)_{CS}$ representation. For $L = 1$
$q\bar{q}$ states, this force is screened by the orbital angular
momentum, but in our case the $\bar{q}$ is in a $S$-wave with
respect to the $4q$ states. The orbital momentum of these $4q$
states is expected to screen the chromo-magnetic force by a factor
which we guess to be $1/2$: it would correspond to the description
of a $\bar{q}$ clustered with one of a $S$-wave $qq$ pair, as
suggested in \cite{HS} many years ago and more recently considered
in \cite{KL1}, but it is worth recalling that the approach used
here is fully consistent with Fermi-Dirac quark statistics.\\ The
lesson to be learned from the successful QCD description of the $q
\bar{q}$ $L = 0$  states is that the chromo-magnetic force is
attractive for small $SU(6)_{CS}$ representations, as dictated by
the general formula,
 with a positive coefficient for the Casimir of the final
$SU(6)_{CS}$ representation.  Therefore in the combination of the
$SU(6)_{CS}$ representations corresponding to the colour triplet
$4q$ $L = 1$ states described in Table 2, with the $\bar{q}$, we
select the smallest possible $SU(6)_{CS}$ representations, to get
the lightest states of the spectrum. These representations are the
$70$ for $210$ and $105$, the $20$ for $105'$ and $\overline{15}$;
in the case of the $105'$ we consider also the $70$, for which the
interaction is also attractive. As previously stated, the $20$ and
$70$ of $SU(6)_{CS}$ contain a colour singlet with spin $3/2$ and
$1/2$, respectively. So for the $Y=2$ pentaquarks with quark
content $uudd\bar{s}$, we may write the phenomenological formula
for the mass spectrum \cite{HS}:
\bea
m=m_0+h\frac{3}{16}( {m_{K^*}-m_K} )
\left[ C_6(p)-C_6(t)-\frac{1}{3}S_p(S_p+1)+\frac{1}{3}
S_t(S_t+1)-\frac{4}{3} \right] + \nonumber \\
\tilde{h}\frac{1}{4}(m_N-m_{\Delta}) \left[ C_6(t)-\frac{1}{3}
S_t(S_t+1)-\frac{26}{3} \right] +a \,\vec{L} \cdot \vec{S}
\label{equation13}
\eea
where $C_6(p)$ and $C_6(t)$ are  Casimir
of $SU(6)_{CS}$ representations, where the pentaquarks and the
tetraquarks are classified, respectively, normalized as
$C_6(35)=6$. The $K^{(*)}$ and the $Y=1$ baryons have been
considered, since they involve the flavours relevant for $Y=2$
pentaquarks. According to our guess we take $h=\frac{1}{2}$ and
$\tilde{h} $ and $a$ from the spectrum of the $70\,L=1$ with $Y=1$
to be $0$ and $40MeV$, respectively. For the diagonalization of
the mass spectrum given by eq.(13) it is useful to know the
Clebsh-Gordon coefficients: 
\bea
&| 20, & (1, S=3/2), S_z= 3/2> = \nonumber \\
& \frac{1}{\sqrt{3}} & \{ \frac{2}{\sqrt{7}} \; | 105'; \, (3, S=2),
S_z =2>_a  \quad
|\bar{6} ; \, (\bar{3}, S= 1/2), S_z = - 1/2>^a \nonumber \\
& \frac{-1}{\sqrt{7}} & |105' ; \, (3, S=2), S_z = 1>_a \quad
|\bar{6} ; \, (\bar{3}, S = 1/2), S_z =  1/2>^a  \nonumber \\
& +\sqrt{\frac{2}{7}} & |105' ; \, (3, S=1), S_z = 1>_a \quad
|\bar{6} ; \, (\bar{3}, S = 1/2), S_z =  1/2>^a  \}
\label{13}
\eea

\bea
&|70, & (1,S = 1/2), S_z = \frac{1}{2} > = \nonumber \\
& \frac{1}{\sqrt{3}} &  \{ \frac{1}{\sqrt{3}} \; | 105' (3, S = 1)
S_z = 1>_a \;
|\bar{6} ; \, (\bar{3}, S= 1/2), S_z = - 1/2>^a \nonumber \\
& \frac{-1}{\sqrt{6}} & | 105' (3, S = 1) S_z = 0>_a \;
|\bar{6} ; \, (\bar{3}, S = 1/2), S_z =  1/2>^a \nonumber \\
& + \frac{1}{\sqrt{2}} & | 105' (3, S = 0) >_a \;
|\bar{6} ; \, (\bar{3}, S = 1/2), S_z =  1/2>^a \} \nonumber \\
\label{14}
\eea

\bea
&|70, & (1,S = 1/2), S_z = \frac{1}{2} > = \nonumber \\
& \frac{1}{\sqrt{3}} &  \{ \frac{1}{\sqrt{2}} \; | 210 (3, S = 1) S_z
= 1>_a \;
|\bar{6} ; \, (\bar{3}, S= 1/2), S_z = - 1/2>^a \nonumber \\
& -\frac{1}{2} & | 210 (3, S = 1) S_z = 0>_a \;
|\bar{6} ; \, (\bar{3}, S = 1/2), S_z =  1/2>^a \nonumber \\
& +\frac{1}{2} & | 210 (3, S = 0) >_a \;
|\bar{6} ; \, (\bar{3}, S = 1/2), S_z =  1/2>^a \} \nonumber \\
\label{15}
\eea

where $a = 1,2,3$ is a colour index to be
saturated to get a colour singlet.

In fact, by neglecting the dependance $S_t$ in Eq.(13), the
l.h.s.'s of eq.'s (14-16) would be eigenvectors of the mass. The
exact consequences of Eq.(13) are reported in Table 3, where the
mass of the $Y=2$ pentaquarks is given for each set of multiplets.
The effects of $SU(3)_F$ breaking will be dealt in a forthcoming
paper \cite{BFHS}.

\begin{center}
\begin{tabular}{|c|c|c|c|}\hline
$SU(6)_{CS} \times S$ & $SU(3)_F {\times} SU(2)_J$ & M ($MeV$) \\
\hline (20*, $\frac{3}{2}$)(105') & (35 + 10 + 27 + 10 + 8 +
$\overline{10}$ + 8 + 8 + 1, $\frac{5}{2}$ + $\frac{3}{2}$ +
$\frac{1}{2}$) & 1640+ 40$(\vec{L}
\cdot \vec{S})$ \\
\hline (70*, $\frac{1}{2}$)(210) & (27 + 10 + 8 + $\overline{10}$
+ 8 + 8 + 1,
$\frac{3}{2}$ + $\frac{1}{2}$) & 1600+ 40$(\vec{L} \cdot \vec{S})$ \\
\hline (70, $\frac{1}{2}$)(105) & (27 + 10 + 8 + 8 + 1,
$\frac{3}{2}$ + $\frac{1}{2}$)
& 1681+ 40$(\vec{L} \cdot \vec{S})$ \\
\hline (20, $\frac{3}{2}$) ($\overline{15}$) & (27 + 10 + 8,
$\frac{5}{2}$ + $\frac{3}{2}$ +
$\frac{1}{2}$ ) & 1755+ 40$(\vec{L} \cdot \vec{S})$\\
\hline (70*, $\frac{1}{2}$)(105') & (35 + 10 + 27 + 10 + 8 +
$\overline{10}$ + 8 + 8 +
1, $\frac{3}{2}$ + $\frac{1}{2}$) & 1742+ 40$(\vec{L} \cdot \vec{S})$\\
\hline (540, $\frac{5}{2})$(105') & (35 + 10 + 27 + 10 + 8 +
$\overline{10}$ + 8 + 8 + 1, $\frac{7}{2}$ + $\frac{5}{2}$ +
$\frac{3}{2}$) & 1854+ 40$(\vec{L}
\cdot \vec{S})$\\
\hline (1134, $\frac{3}{2}$)(210) & (27 + 10 + 8 + $\overline{10}$
+ 8 + 8 + 1, $\frac{5}{2}$ + $\frac{3}{2}$ + $\frac{1}{2}$) & 1866
+40$(\vec{L}
\cdot \vec{S})$ \\
\hline (560, $\frac{3}{2}$)(105) & (27 + 10 + 8 + 8 + 1,
$\frac{5}{2}$ + $\frac{3}{2}$ + $\frac{1}{2}$)
& 1866 +40$(\vec{L} \cdot \vec{S})$ \\
\hline (540*, $\frac{3}{2}$)(105') & (35 + 10 + 27 + 10 + 8 +
$\overline{10}$ + 8 + 8 + 1, $\frac{5}{2}$ + $\frac{3}{2}$ +
$\frac{1}{2}$) & 1882+ 40$(\vec{L}
\cdot \vec{S})$\\
\hline (1134*, $\frac{1}{2}$)(210) & (27 + 10 + 8 +
$\overline{10}$ + 8 + 8
+ 1, $\frac{3}{2}$ + $\frac{1}{2}$) & 1885 +40$(\vec{L} \cdot \vec{S})$ \\
\hline (540*, $\frac{1}{2}$)(105') & (35 + 10 + 27 + 10 + 8 +
$\overline{10}$ + 8 + 8 +
1, $\frac{3}{2}$ + $\frac{1}{2}$) & 1885+ 40$(\vec{L} \cdot \vec{S})$\\
\hline (70, $\frac{1}{2}$) ($\overline{15}$) & (27 + 10 + 8,
$\frac{3}{2}$ + $\frac{1}{2}$ ) & 1903+ 40$(\vec{L} \cdot \vec{S})$\\
\hline
\end{tabular}
\end{center}
\vspace*{0.1cm}
\begin{center}
Table 3: Mass spectrum of the positive parity pentaquarks built
with $4q$ with $L=1$ and a $\bar{q}$ in S-wave.
The * is put to remind of a mixing between the $ SU(6)_{CS}$
representations and the transformation properties of the $4q$
state have been written in brackets, since they are relevant for
the $SU(6)_{CS}$ selection rule, which is shown in the following
lines.
\end{center}
We have a \underline{ selection rule} in $SU(6)_{CS}$
analogous to the one found in the framework of $SU(6)_{FS}$
\cite{SR}. This new selection rule comes from the fact that the
$(10, \frac{3}{2}^+)$ and the $(8, \frac{1}{2}^+)$ transform as
the 20 and 70 $SU(6)_{CS}$ representations, respectively. The
states of the $210$ and of the $105$ $SU(6)_{CS}$ cannot decay
into meson decuplet states, while the states of the
$\overline{15}$ cannot decay into meson octet states. Let us
emphasize that in this case the selection rule is not affected by
$SU(3)_F$ and $SU(2)_I$ breaking. In conclusion, only  the states,
with their $4q$ transforming as the $105'$ of $SU(6)_{CS}$ may be
found by looking
for decuplet-meson final states in octet-meson reactions.\\
The lightest multiplets are the $J^P=\frac{1}{2}^+$ states  built
by combining the $S=\frac{3}{2}$ state approximately given by
Eq.(14) with the orbital momentum $L=1$. In the $\bar{10}$ we
propose to classify the $\Theta^+$ and the $\Xi^{I=\frac{3}{2}}$
particles. For this reason we have fixed $m_0$ in Table 3 to
reproduce the mass of $\Theta^+$. In fact that $\bar{10}$ has
reduced couplings to MB final states as, which come by the product
of three factors.\\
In fact the $(\bar{6}, S = 2)$ multiplet is in the $105$ of
$SU(6)_{FS}$ and obeys the selection rule preventing the decay
into MB (i.e. meson-baryon) state. The $(\bar{6}, S = 1)$ of the
$SU(6)_{CS}$ $105'$, for which the exact computation upgrades the
factor $\sqrt{\frac{2}{7}}$ to $\simeq \sqrt{\frac{3}{8}}$, is an
equal mixture of the $SU(6)_{FS}$ $210$ and $105'$, which implies
a reduction factor of $\frac{1}{\sqrt{2}}$. Another reduction
factor $\frac{1}{\sqrt{2}}$ comes by considering the SL tensor
product $\frac{3}{2}$ x $1$, which gives the total angular
momentum $\frac{1}{2}$ of the pentaquark. The $S_z = \frac{3}{2}$
state appears with a factor $\frac{1}{\sqrt{2}}$ in that
pentaquark state and the component coming from the S=1 state,
which is the one, for which the decay into MB is allowed, has
$S^q_z\,=\,1$ and $S^{\bar{q}}_z\,=\,\frac{1}{2}$. Therefore the
$\bar{q}$ to form a meson should take the quark with opposite
$S_z$, leaving the remaining quarks with $S_z\,=\,\frac{3}{2}$
with the wave function orthogonal to the baryon octet ( the decay
into the meson plus decuplet of a particle in a  $\bar{10}$ of
$SU(3)_F$ violates $SU(3)_F$ symmetry). In conclusion we predict a
global reduction factor of $\frac{3}{32}$ for the width of the
lightest decuplet state according to Eq.(13), which is welcome to
explain the narrowness of $\Theta^+$ and $\Xi^{I=\frac{3}{2}}$
states. In the case of the $\Xi^{--}$, the narrow width of which
is still more surprising due to the large phase space available
for decay into $\Xi^- \pi^-$, there may be a distructive
interference between the contribution of the $210$ $SU(6)_{FS}$
state and the flavour-violating
contribution, which one expects for the states of the $105 + 105'$.\\
  The preliminary result, announced at the
recent Jefferson Laboratory pentaquark conference, of a $\Xi^*
(1530) \pi$ resonance found by NA49 at a mass near to the
$\Xi^{--}$ (and $\Xi^0$) found in $\Xi^- \pi^- (\pi^+)$ final
states with a comparable number of events above the background
would be a striking confirmation of our proposal. Indeed, we
expect amplitudes of the same order for the $SU(3)$ allowed
$\Xi^{I=3/2}$ decays into $\Xi \pi$ and for the $SU(3)$ forbidden
$\Xi^{I=3/2} \to \Xi^*\pi$ decay (the product $10 {\times} 8 = 35
+ 27 + 10 + 8 $ does not contain a $\overline{10}$ representation)
\footnote{We thank M. Karliner for bringing this fact to our
attention during his seminar at CERN.} .\\
We can develop for the $3$'s of $SU(3)_F$, transforming as
the $105'$ of $SU(6)_{CS}$, the same considerations  as for the
$\bar{6}$'s, leading to the conclusion of  the existence
of the multiplet of states $(\overline{10} + 8 + 8 + 1, 5/2^+ +
3/2^+ + 1/2^+)$ weakly coupled to the $MB$ channels. The
$\overline{10} \; 1/2^+$ is the state, we have proposed for the
classification of the $\Theta^+$ and $\Xi^{I=3/2}$ discovered
pentaquarks. One might think that the three quarks left in a
parity negative state by the S-wave emitted meson could produce a
$q\bar{q}$ pair to give a $MB \pi$ final state. The $\Theta^+$
state is below the threshold for producing a $KN\pi$ state, but
for $\Xi^{I=3/2}$ it would be
worth looking for $\Xi \pi \pi$ final states.\\
The low couplings
to MB final states of the $(8 + 8 + 1, 1/2^+)$ multiplets, as
well as those of the remaining states of the $\overline{10}$,
which are classified in the same multiplet of $\Theta^+$ and
$\Xi^{I=3/2}$, explain why they succeeded up to now in escaping
observation. They have the same $I,Y$ quantum numbers of $3q$
states strongly coupled to the meson-baryon channels, and the few
events for which they are responsible are hidden by the
overwhelming number of events due to $3q$ resonant states.  Only
very accurate experiments, encouraged by the knowledge of their
existence and made easier by some prejudice on the values of their
masses, would reveal them. The other light multiplets built with
the $(15' + 15,S=2)$ tetraquarks of the $105'$ representation of
$SU(6)_{CS}$ have not reduced couplings to MB final states, since
they transform as the 126 and 210 representations of $SU(6)_{FS}$,
for which the flavour selection rule discovered in \cite{SR} does
not apply. More in general all the states with large components
along the $210$ and $126$ $L =1$ $SU(6)_{FS}$ multiplets are
allowed to decay into $BM$ and not expected to be narrow. The
Roper resonance may be classified as one of these states. In
particular the $35$ and $27$ of $SU(3)_F$ contain $Y = 2, I = 2$
(to be looked for in the $N^{*++} K^+$ final state)
and $I = 1$ states, respectively.\\
An inspection to the mass spectrum described in Table 3 shows that
all the states lay below $2\,GeV$, below the threshold to decay
into $K$ negative parity baryon states, to which the pentaquark
particles here described are expected to have large couplings.
However we expect some of their $SU(3)_F$ partners to be above the
threshold to decay into $\pi$ negative parity baryon states.

We may apply Eq.(13) to the study of the spectrum of the negative
parity states constructed with $4q$ and a $\bar{q}$ in a $S$-wave,
we have mentioned at the beginning of this letter. We expect  a
spectrum with the $3, \bar{6}, 15$ and $15'$ representations with
increasing mass, since they transform as the $210, 105, 105'$ and
$\overline{15}$ representations of $SU(6)_{CS}$, respectively.
From Eq.(13) with a different value of the central mass, we call
for them $\tilde{m_0}$, and $h = \tilde{h} = 1$, since we do not
expect the chromo-magnetic to be screened, we find the spectrum
reported in Table 4.
\begin{center}
\begin{tabular}{|c|c|c|}\hline
$SU(6)_{CS}$             & $SU(3)_F \times SU(2)_{J=S}$                   &    M ($MeV$)               \\
\hline 70*(210)          &(8 + 1, $\frac{1}{2})$                          & $\simeq\tilde{m_0}-669$    \\
\hline 70(105)           & ($\overline{10}$ + 8, $\frac{1}{2}$ )          & $\tilde{m_0}-345$       \\
\hline 20*(105')         & (27 + 10 + 8 , $\frac{3}{2}$ )                 & $\tilde{m_0}-222$       \\
\hline 1134(210)         & (8 + 1, $\frac{3}{2}$) &
$> \tilde{m_0}-123$ \\
\hline 70*(105')         & ( 27 + 10 + 8 , $\frac{1}{2}$ )                       & $\tilde{m_0}-107$ \\
\hline 560(105)          & ($\overline{10}$ + 8, $\frac{3}{2}$)                  & $\tilde{m_0}+25$\\
\hline 540*(105')         & ( 27 + 10 + 8 , $\frac{1}{2}$)                        & $\tilde{m_0}+205$\\
\hline 540(105')         & (27 + 10 + 8 , $\frac{5}{2}$ )         & $\tilde{m_0}+247$  \\
\hline 540*(105')         & (27 + 10 + 8 , $\frac{3}{2}$ )         & $\tilde{m_0}+247$  \\
\hline 20($\overline{15})$& (35 + 10 , $\frac{3}{2}$)                           & $\tilde{m_0}+247$ \\
\hline 1134*(210)           & (8 + 1, $\frac{1}{2})$  & $> \tilde{m_0}-322$\\
\hline $70(\overline{15})$      & (35 + 10 , $\frac{1}{2}$)                             & $\tilde{m_0}+543$\\
\hline
\end{tabular}
\end{center}
\vspace*{0.1cm}
\begin{center}
Table 4: Spectrum of negative parity states built with $4q$ and a
$\bar{q}$ in S-wave. The notations are as in Table 3 and the mass
of the $Y=2$ states are reported with the only exception of the $
8 + 1 $ multiplets with $4q$ transforming as the $210$ of
$SU(6)_{CS}$, for which the mass reported is the one of the $Y=0$
$I=1$ state ( with quark content $uuds\bar{d}$ ). The fact that
there is a strange $q$, instead of a $\bar{q}$ will make the
numbers written in Table 4 for these states affected by $SU(3)_F$
breaking, but the effect should be less relevant for the lightest
state, for which the two relevant terms in Eq.(13) give comparable
contributions, than for the other two multiplets: this motivates
the different mathematical symbols present in Table 4.
\end{center}
The spectrum of the positive parity states built with $4q$ in
S-wave and a $\bar{q}$ with $ L = 1 $ with respect to them may be
found by Eq.(13), by taking $ h = 0 $ , $ \tilde{h} = 1$ and $ a =
40
MeV$.\\
By applying Eq.(13) to $qq\bar{q}\bar{q}$ meson states with $ h =
\tilde{h} = 1$, modified to keep into account that the
chromo-magnetic interaction is slightly stronger for $ud$ quarks
than for $s$ ( $m_{\rho} - m_{\pi} = \frac{4}{3} (m_{K^*} -m_K)$ ),
one would obtain the intriguing result that the lightest state,
with a contribution of the chromo-magnetic interaction $\simeq - 1
GeV$, is a $ I = 0 $ state with quark content $ud\bar{u}\bar{d}$,
which transforms as a singlet of $SU(6)_{CS}$, to be identified
with the $f^0(600)$ $0^+$ state. Several hundreds of $MeV$ above
that state one predicts a ($I = 1 + 0$, $0^+$) $qs\bar{q}\bar{s}$
multiplet to be identified with the $f^0(980)$ and $a_0(980)$
$0^+$ states, for which the $qs\bar{q}\bar{s}$ content has been
already proposed \cite{WI}. \\
As suggested in \cite{KL2}, we should look for narrow
$\bar{D}N$ and $BN$ resonances that we would obtain by
substituting, in $\Theta^+$, $\bar{s}$ with $\bar{c}$ or
$\bar{b}$, respectively. The states with $4q$, one of which is
heavy, forbidden by the $SU(6)_{FS}$ selection rule to decay into
a baryon heavy meson state (let us think of a $N^{*P=-1} D$ bound
state) are not expected to be narrow, because for them the channel
meson heavy baryon is opened and the $SU(6)_{FS}$ selection rule
is not effective by the large $SU(4)_F$ breaking.

{\bf Acknowledgements}\\

We are indebted to H. H\"{o}gaasen for valuable
discussions.

\end{document}